\documentclass[12pt,preprint]{aastex}
\usepackage{epsf}
\usepackage{lscape}
\usepackage{graphicx}
\usepackage{natbib}
\title {MAPPING THE UNIVERSE: THE 2010 RUSSELL LECTURE \footnote {This paper preserves the substance and style of Margaret Geller's 2010
Russell Lecture presented at the May 2011 Boston AAS Meeting.}}

\author {Margaret J. Geller} 
\affil{Smithsonian Astrophysical Observatory,
\\ 60 Garden St., Cambridge, MA 02138}
\email{mgeller@cfa.harvard.edu}
\author{Antonaldo Diaferio}
\affil{Dipartimento di Fisica Generale 'Amedeo Avogadro', 
\\Universit\`a degli Studi di Torino, via P. Giuria 1, 10125 Torino, Italy}
\affil{ INFN, Sezione di Torino, via P. Giuria 1, 
\\10125 Torino, Italy}
\affil{ Harvard-Smithsonian Center for Astrophysics,
\\60 Garden Street, Cambridge, MA 02138}
\email{adiaferio@cfa.harvard.edu}
\author {Michael J. Kurtz} 
\affil{Smithsonian Astrophysical Observatory,
\\ 60 Garden St., Cambridge, MA 02138}
\email{mkurtz@cfa.harvard.edu}

\begin{document}

\begin{abstract}
Redshift surveys are a powerful tool of modern cosmology. We discuss two aspects of their power to map the distribution of mass and light in the universe: (1) measuring the mass distribution extending into the infall regions of rich clusters and (2) applying  deep redshift surveys to the selection of clusters of galaxies and to the identification of very large structures
(Great Walls). We preview the HectoMAP project, a redshift survey with median redshift 
$z = 0.34$ covering 50 square degrees to r= 21. We emphasize the importance and power of spectroscopy for exploring and understanding the nature and evolution of structure in the universe.
\end {abstract}

\section {Introduction}

In the fall of my senior year (1969) as a Berkeley physics undergraduate, Charles Kittel gave me some startling advice. He, a pioneer of solid state physics, told me that I should not enter that field because it was mature ...``done.'' He told me that I should choose a field that would still be ``new'' when I was ten years past my PhD. and a ``mature'' scientist.
He suggested that I think about  astrophysics and biophysics and pressed me to apply to Princeton to do astrophysics. I did apply to Princeton and was admitted to the physics department. Within my first year of graduate school, I asked Jim Peebles to supervise my research ... and he agreed.

It was a very special and wonderful time to be starting out in astrophysics, particularly in the Princeton physics department. Dave Wilkinson and his group were leading the study of the relatively recently discovered cosmic microwave background. Jim Peebles was thinking about the distribution of galaxies in the universe and he started me on the track of finding ways to extract physical constraints from catalogs of galaxies with redshifts. 
The total store of redshifts was shockingly small by today's standards.  In the second paper Peebles and I wrote together, we used a catalog of {\bf 527} redshifts along with an n-body simulation to make one of the first statistical estimates of the masses  of galaxies (Geller \& Peebles 1973). 

\begin{figure}[htb]
\centerline{\includegraphics[width=4.0in, angle = -90]{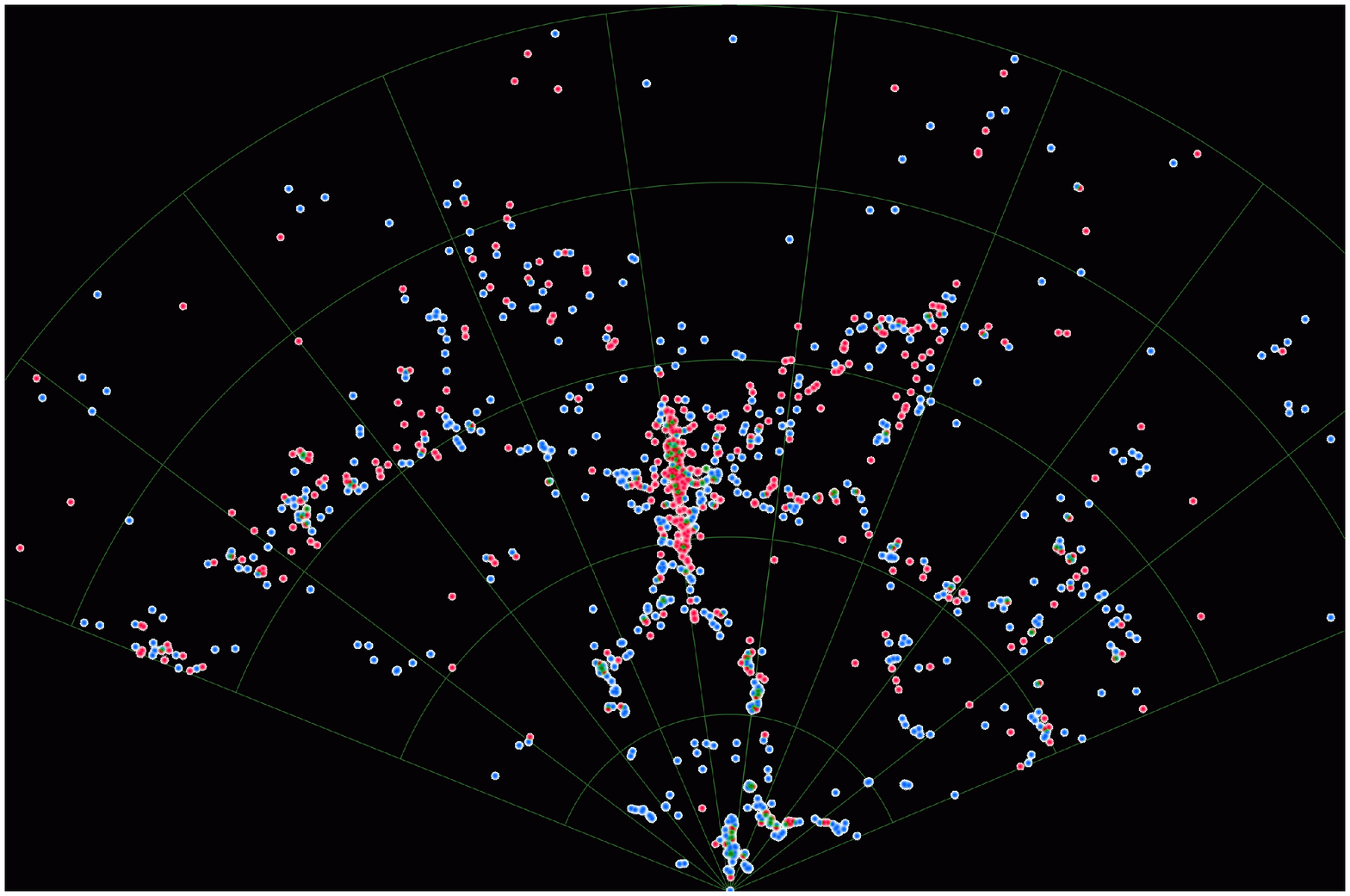}}
\vskip 5ex
\caption{Slice of the universe after de Lapparent, Geller \& Huchra (1986). The points represent individual galaxies; blue are late-type and red are early-type (Huchra et al. 1990).
The radial grid marks redshift intervals $\Delta{z} = 0.01$; the azimuthal coordinate is right ascension and the grid runs from 8 to 17 hours in 1 hour intervals. Note the cut through the
Great Wall at $z \sim 0.02-0.03$ and the prominent ``finger'' of the Coma cluster in the center of the map.
\label{fig:firstslice}}

\end{figure}

As a student I could not imagine how rapidly our ability to map the universe would change. I never would have predicted that in 1985 Valerie de Lapparent, John Huchra, and I would measure
redshifts for $\sim$1100 galaxies in a slice of the universe and that the stunning pattern they revealed would change the general perception of the way galaxies are arranged on large scales
(de Lapparent, Geller \& Huchra 1986).  Figure
\ref {fig:firstslice} shows the now iconic stick figure pattern in our slice. The torso of the stick figure is the
``finger'' of the Coma cluster. The band of galaxies running all the way across the survey  
is a cut through the Great Wall (Geller \& Huchra 1989). The sharply outlined voids surrounded or nearly surrounded by thin filaments and sheets containing galaxies are the hallmarks of what we now call the ``cosmic web.''

During the last twenty years wide-field multi-object spectrographs have revolutionized our ability to map the universe. The number of redshift measurements has increased exponentially;
the NASA/IPAC Extragalactic Database includes $\sim 2$ million redshifts. Ambitious surveys including the Sloan Digital Sky Survey (Abazajian et al. 2009), 6DF (Jones et al. 2009), and 2DF (Colless et al. 2001) are major contributors to this wealth of data . In 2000, I calculated that if
technology continued to improve as it had up to that point, we would have a redshift for every galaxy in the visible universe by the year 2100. Perhaps that will really happen.

Rather than review the many remarkably successful projects that have changed the field in the last years, I want to focus on two projects we have carried out with the multi-object spectrograph, Hectospec, on the MMT. These two projects grow out of features revealed in our first slice of  the universe: (1) the hint of a trumpet-shaped infall pattern around the Coma cluster and (2) the Great Wall. 

The work I discuss here has not been published previously. I thus include Antonaldo Diaferio and Michael Kurtz as co-authors. This inclusion is a small thank you for their support and for the joy of working with them on these projects and many others. The Introduction and Concluding Remarks are my voice alone. Section \ref{caustics} is co-authored by Antonaldo Diaferio and Section \ref{HectoMAP} is co-authored by Michael Kurtz.

Section \ref{caustics} discusses redshift surveys of the infall regions of clusters of galaxies. It includes a tutorial movie of a simulation of the evolution of a cluster in real and redshift space. We review the idea of the caustic method for estimating the mass within the infall region and we briefly discuss some of the results of applying this technique to data from the SDSS along with new data from Hectospec.

Section \ref{HectoMAP} announces HectoMAP, a redshift survey with a median depth $ z = 0.34$. The survey covers a 50 square degree strip of the northern sky and will eventually include 60,000 redshifts. The goals of the survey include the study of clusters of galaxies and their environment at moderate redshift.  We preview the survey and discuss the suggestion of Great Walls at greater and greater redshift.

\section{Clusters of Galaxies in Redshift Space}
\label{caustics}

Clusters of galaxies are a cornerstone of modern cosmology. Zwicky's  first  application of the
virial theorem to a few redshifts in the Coma cluster showed that clusters must contain dark matter (Zwicky 1933). This pioneering paper set the stage for the use of kinematic measures as a route to understanding the matter distribution in the universe. Today wide-ranging studies of clusters of galaxies reaching from the nearby universe to large redshift provide constraints on the growth of structure in the
universe and on the cosmological parameters (e.g. Haiman, Mohr \& Holder 2001; Voit 2005; Cunha, Huterer \& Frieman 2009; Pierre et al. 2011).

Now a host of techniques are available to probe the matter distribution 
within clusters of galaxies. Different techniques are applicable at different radii.
The fiducial radii R$_{500}$ and R$_{200}$ are the radii enclosing a matter density
500 and 200 times, respectively, the critical density.
Within R$_{500}$, x-ray observations and strong lensing provide important constraints. Galaxy dynamics and scaling relations extend the reach of our knowledge to R$_{200}$. Generally 
R$_{200}$ is comparable with the extent of the virialized (relaxed) core of a cluster. Analyses of x-ray observations and dynamical calculations on these scales make a variety of equilibrium and symmetry assumptions. Generally, agreement among the various mass estimation techniques on this scale is impressive. Although there are still puzzles about clusters and their evolution, their central regions are reasonably well-studied over a wide redshift range. 

Many fewer observational studies have addressed the infall region that marks the transition between the cluster core and the surrounding large-scale structure. At least in part, this
inattention reflects the observational challenges of observing these larger, less dense regions.
Now with wide-field spectroscopic instruments like the Hectospec on the MMT
(Fabricant et al. 1998; Fabricant et al. 2005), it is possible to acquire dense samples of these
fascinating regions that lie between R$_{200}$ and R$_{turn}$, the radius of the shell of material just turning around from the Hubble flow at redshift $z$ (Gunn \& Gott 1972; Kaiser 1987; Regos \& Geller 1989).

The infall region is  a route to understanding the growth rate of clusters, their ultimate masses, and the relationship between galaxy and cluster evolution (Diaferio \& Geller 1997; Ellingson et al. 2001; 
Busha et al. 2005; Rines et al. 2005; Tran et al. 2005). 
On the scale of the infall region, there are only two techniques to probe the matter distribution, weak lensing (e.g. Lemze et al. 2009; Umetsu et al. 2011) and a kinematic technique called the caustic method (Diaferio \& Geller 1997; Diaferio 1999; Serra et al. 2011). Neither of these methods depends on the dynamical state of the system and both apply at all clustrocentric radii (Diaferio, Geller \&
Rines 2005).

Of course, for nearly all clusters, we can observe them only in redshift (phase) space. Kaiser
(1987) was the first to understand how spherical infall appears in redshift space. In his elegant paper (Kaiser 1987), he shows (his Figure 5) the now widely recognized trumpet-shaped pattern
that characterizes the appearance of a cluster in redshift space. The central, virialized region appears as an extended finger pointing along the line-of-sight toward the observer. This elongation is a simple consequence of the fact that the line-of-sight component of the 
velocities of galaxies relative to one another within the virialized region are larger than the Hubble flow across the region. At the effective outer radius of the cluster, R$_{turn}$, the
infall velocity just cancels the Hubble flow. Thus the shell just turning around appears as a line at the cluster mean velocity in redshift space. Infalling shells at radii between R$_{turn}$ and  R$_{200}$ are successively more and more elongated along the line-of-sight producing the trumpet shape. In the simple spherical infall model, the outline of the trumpet is a true caustic (a line of infinite density in phase space).

At about the same time that Kaiser wrote his paper, there was an increasing awareness of the complexity of cluster evolution. A substantial fraction of clusters observed with the Einstein
X-ray Observatory showed substructure in their surface brightness distribution (Forman et al. 1981; Jones \& Forman 1984). Corresponding structure  appeared in the distribution of galaxy counts on the sky (Geller \& Beers 1982) and in the line-of-sight velocity distribution for cluster members (Dressler \& Shectman 1988). All of these observations provided strong support for the now standard hierarchical picture of structure formation where clusters are built up by the coalescence of smaller galaxy groups. They also show that clusters are still in the process of formation at the current epoch. 

\begin{figure}[htb]
\centerline{\includegraphics[width=6.0in]{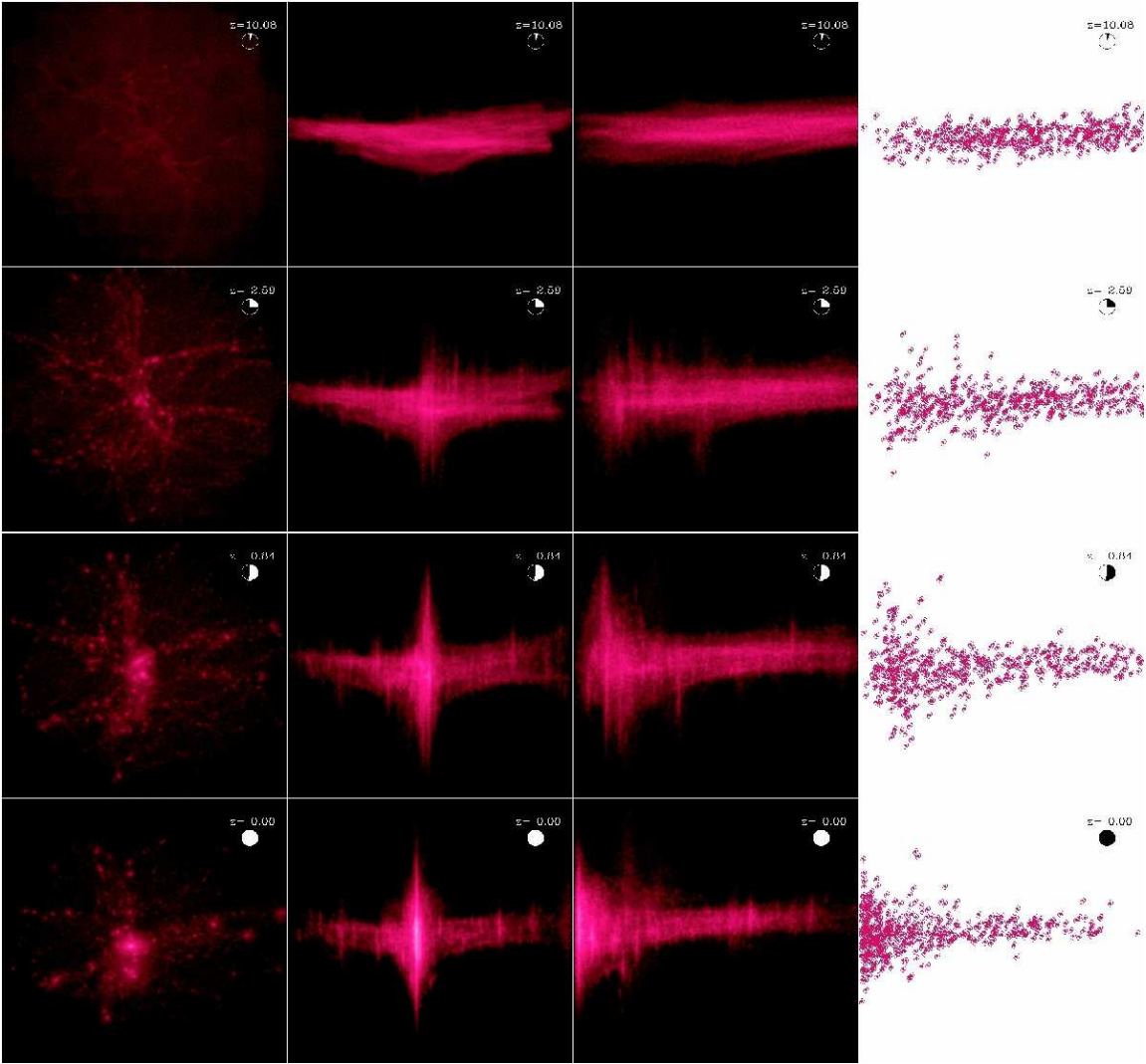}}
\vskip 5ex
\caption{Montage of images from an n-body simulation of the development of a cluster
with M$_{200} = 6.28\times 10^{14}h^{-1}$M$_\odot$. The first column shows the distribution of dark matter in the $x-y$ plane of configuration space.
The spatial coordinates range from  -15 to +15 Mpc/$h$ in the rest-frame of the cluster at each redshift. The second column shows a section of phase space in the same frames:
the $x$ spatial coordinate is  on the abscissa  and ranges from -15 to +15 Mpc/h;
the peculiar velocity along the $z$ spatial coordinate is shown on the ordinate 
and ranges from -4000 to +4000 km s$^{-1}$. The third column shows the
redshift diagram: the x-axis is the projected distance in the $x-y$ plane of configuration space from the cluster center defined at redshift $z=0$; the axis extends to 15Mpc/h. The vertical axis is the rest-frame line-of-sight velocity and  extends from -4000 to +4000 km s$^{-1}$. The fourth column shows the effect of sparse sampling: it shows 500 randomly sampled particles in the redshift diagrams of the third column. The number in the upper right of each panel is the redshift; the clock shows the fraction of cosmic time elapsed since the start of the simulation at redshift $z = 20$. The on-line journal article includes a corresponding 4-panel video; the bottom row of this figure shows the final frames of the video.
\label{fig:simulation}}

\end{figure}

Increasingly sophisticated n-body simulations have provided a guide to the complex evolution of clusters of galaxies. Figure \ref{fig:simulation} shows snapshots of the evolution of the cluster in
configuration space (left), in redshift space (central two columns), and as traced by ``galaxies'' (right); an accompanying 4-panel movie shows the full evolution of the cluster. 

The simulation models the formation of a galaxy cluster in a $\Lambda$CDM cosmological
model. We use a multi-mass technique with vacuum boundary conditions (Tormen \& Bertschinger 1996; Springel et al. 2001).  The simulation
was run with GADGET (Springel, Yoshida \& White 2001) and contains $1.4\times 10^5$ particles in the central high-resolution region out of a total  of the $2.8\times 10^5$ particles. The particles in the high-resolution region have $1.1\times 10^{10} h^{-1} M_\odot$
where the Hubble constant is 100$h$ km s$^{-1}$Mpc$^{-1}$. The more massive particles in the external, low-resolution regions mimic the tidal field of the large-scale structure. The cluster at redshift $z=0$ has $M_{200}=6.28\times 10^{14} h^{-1} M_\odot$ and $r_{200}=1.39 h^{-1}$ Mpc.

In the beginning
there are small irregularities in the dark matter density (first row, Figure \ref{fig:simulation}). As the cluster evolves gravity amplifies these small irregularities and structure grows. As the cluster evolves, condensations into the dark matter distribution (groups) flow along the surrounding filaments (walls) into the
central mass concentration (second row, Figure \ref{fig:simulation}). In this  cluster there is a merger of major subclumps at a redshift of $\sim 0.8$ (third row, Figure \ref{fig:simulation}).

Of course, we can never observe the evolution in 3D configuration space. Nonetheless it is rare to see these simulations displayed in redshift space. Figure \ref{fig:simulation} and the associated video show the evolution in redshift space (middle two columns). In the second column we preserve the spatial coordinate along the x-axis and plot the rest-frame
peculiar velocity along the vertical axis. 
As the cluster evolves fingers corresponding to groups of
various masses appear; they flow toward the central mass concentration (we define the cluster
center in redshift space at redshift $z = 0$ and trace the position back through the simulation). Finally the characteristic trumpet-shape pattern appears with the strongly elongated finger in the virialized cluster center. The amplitude of the elongation decreases until it matches the
line-of-sight velocity dispersion of the surrounding structure. It is fascinating that as Kaiser predicted from the spherical infall model, the infall pattern surrounding the cluster appears as two trumpet horns stuck together. Here, however, detailed analyses demonstrate that the boundary of the cluster is not a caustic. In fact the caustics are, in effect, smeared out by the peculiar motions within the infalling structures. Even so, the underlying pattern remains. 

Diaferio and Geller (1997) and Diaferio (1999) used earlier simulations to explore approaches to the
determination of the mass distribution in the infall region where, obviously, the equilibrium assumptions used to analyze cluster cores do not apply. They showed that the amplitude of the
pattern in Figure \ref{fig:simulation} is a measure of the escape velocity from the cluster. They developed a method of identifying the steep change in phase density that defines the infall pattern 
and called their mass estimation technique the ``caustic'' technique in recognition of Kaiser's early work. Their approach enables measurement of the cluster mass profile to large radius.

Figure \ref{fig:simulation} provides a guide to the translation from the simulations to analysis of the data. The third column shows snapshots of the evolution of the cluster in redshift space, but in contrast with the snapshots in the second column, the cluster center is on the left edge of the plot. The abscissa is the projected comoving distance from the cluster center and the ordinate is the same as in the second
column, the line-of-sight rest frame peculiar velocity relative to the cluster center. In essence this plot shows an azimuthal sum over the representation in the second column. Of course, we can never observe the dark matter distribution; we sample it by measuring redshifts of individual galaxies in the cluster. 

Figure \ref{fig:simulation}  (left column) also shows what happens when we sample velocity distribution with 
500 ``galaxies'', about the number of objects we could reasonably observe in a single system.
The ``galaxies'' in Figure \ref{fig:simulation} are a randomly chosen subset of the dark matter particles; i.e we assume the galaxies are unbiased tracers of the dark matter distribution
(e.g. Diaferio 1999; Diaferio et al. 1999). This sampling makes it painfully obvious that we are observationally blind to most of the complex structure in the infall region; there just are not enough galaxies to sample the dark matter distribution densely enough to reveal all of the subsystems that come together to form the cluster.

\begin{figure}[htb]
\centerline{\includegraphics[width=6.5in]{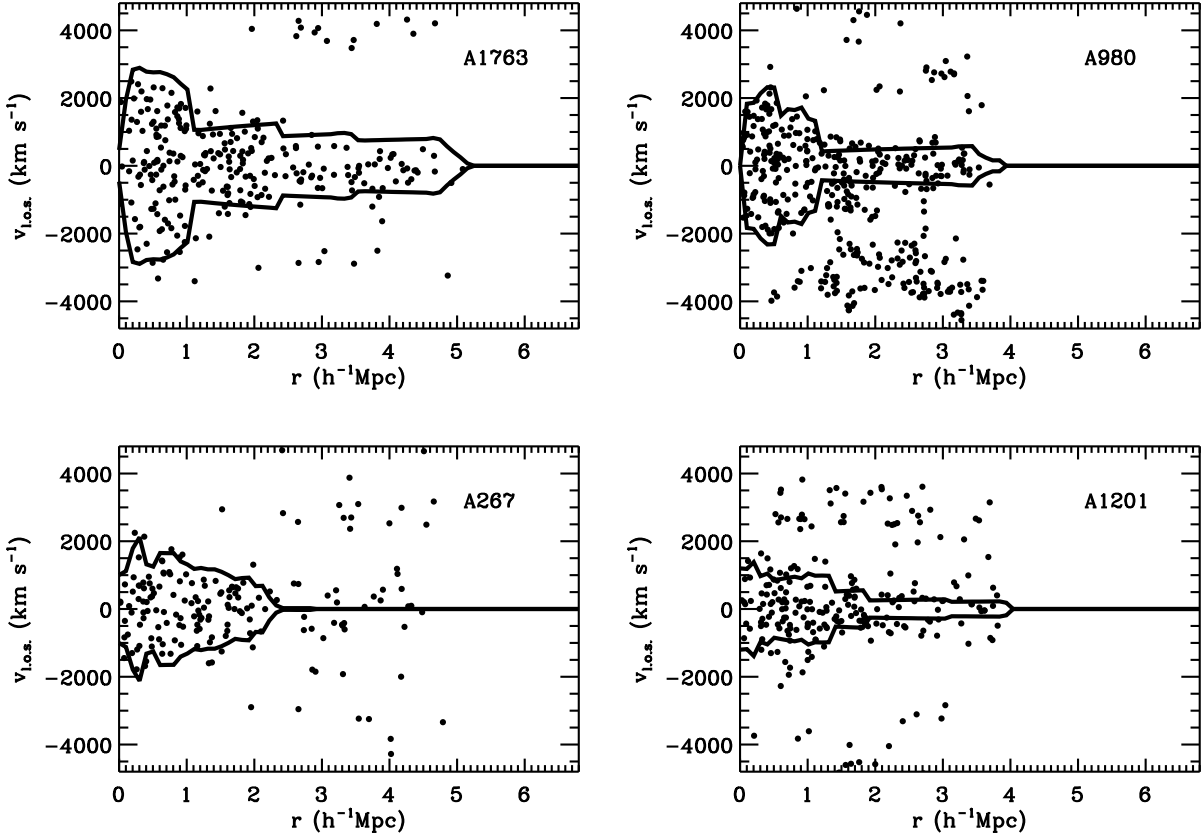}}
\vskip 5ex
\caption{Redshift space diagrams for four Abell clusters. The solid curves mark the
cluster boundaries in redshift space. For A1763 there are 220
galaxies within the ``caustics'' and  $M_{200}=1.35\pm 0.28 \times 10^{15} h^{-1} M_\odot$.
For A980 there are 196 galaxies inside the ``caustics'' and  $M_{200}=5.06\pm 0.79 \times 10^{14} h^{-1} M_\odot$. For the lower mass cluster 
A267 there are 107 galaxies and  $M_{200}=2.83\pm 0.12 \times 10^{14} h^{-1} M_\odot$ and for
A1201 there are 112 galaxies with $M_{200}=9.51\pm 0.57 \times 10^{13} h^{-1} M_\odot$. Note how the infall pattern shrinks as the mass decreases. The cluster mean redshifts are 0.232, 0.155, 0.229 and 0.167, respectively (the zero point on the abscissa corresponds to this mean in each case).
\label{fig:hecs}}

\end{figure}

In spite of these fundamental limitations, application of the caustic technique to data on many clusters
has provided some interesting, new insights. Figure \ref{fig:hecs} shows data for 4 
from the HeCS (Hectospec Cluster Survey) sample of Rines et al. (2011, in preparation). The clusters range in M$_{200}$
from $9.5 \times 10^{13}$M$_\odot$ to $1.4 \times 10^{15}$M$_\odot$ .  For the most massive system in Figure \ref{fig:hecs}, $\sim$ 200 galaxies define the infall pattern; for the least massive there are $\sim$100 galaxies.  The observations match the qualitative appearance of the infall patterns in the simulations. 

Redshift surveys of clusters extending to large radius reveal that the trumpet-shaped
patters are ubiquitous (Rines et al. 2003; Rines \& Diaferio 2006). The majority of massive clusters are well-separated from foreground and background structures in redshift space. Rines \& Diaferio (2006) show that masses computed from the caustic technique
within the virial radius correspond well with those computed from the x-ray and from other dynamical techniques. Diaferio, Geller \& Rines (2005) show that mass profiles derived from weak lensing agree well with those derived from the caustic technique.

One of the most interesting results of the application of the caustic technique is the evaluation of the mass contained within the infall region. Rines \& Diaferio (2006) show that the mass within R$_{turn}$ is 2.19$\pm$0.18 times the mass inside R$_{200}$, in excellent agreement with theoretical predictions of the ultimate masses of clusters (Busha et al. 2005).

We can explore the infall regions of clusters efficiently now because of the power of wide-field
spectrographs like Hectospec. But, when Hectospec was just ``an idea'', the main scientific motivation was to map the large-scale structure of the universe to greater redshift (Geller 1994). Clusters of galaxies are markers of the highest density regions of the universe, but they are a poor second to seeing the entire grand pattern
of the ``cosmic web.'' The first large-area survey with Hectospec is now  underway. We call it HectoMAP.

\section {HectoMAP: Greater and Greater Walls}
\label{HectoMAP}

In 1989,
the Great Wall was the largest structure known in the universe (Geller \& Huchra 1989). It still seems
remarkable that the largest structure  was as big as it could be to fit within the survey boundaries. The Sloan Digital Sky Survey contains the Sloan Great Wall. Estimates suggest that the Sloan Great Wall is only 80\% greater in extent than the CfA Great Wall (Gott et al. 2005) even though the Sloan redshift survey is more than three times as deep as the CfA slices. The Sloan Great Wall is, nonetheless, a potential challenge to our understanding of the development of the cosmic web from Gaussian initial conditions (Sheth \& Diaferio 2011).

The obvious question is whether the Sloan uncovered the biggest structure; after all, the Sloan Great Wall could be bigger and still fit in the survey. One of the goals of our new survey, HectoMAP, is to begin to approach this question by carrying out a deep dense redshift survey
over an area large enough to detect ``greater walls.''

HectoMAP is a redshift survey of red galaxies ($g - r > 1$ and $r - i > 0.5$) with SDSS r$_{petro} < 21$ and r$_{fiber} < 22$
covering a 50 square degree region of the sky with $200^\circ < \alpha_{2000} < 250^\circ$
and $42.5^\circ < \delta_{2000} < 44^\circ$. We select galaxies from the SDSS. The complete survey will include $\sim$ 60,000 redshifts. To date the survey includes $\sim$42,000 redshifts.

We began observations in the HectoMAP strip in 2009 and they will continue at least through the  spring of 2012. 
We acquired spectra for the objects with the Hectospec 
(Fabricant et al. 1998, 2005), a 300-fiber robotic instrument mounted on the MMT. The Hectospec observation
planning software (Roll et al. 1998) 
enables  efficient acquisition of a pre-selected sample of galaxies. The software enables assignment of priorities as a function of galaxy properties.

The spectra cover the wavelength range 3500 --- 10,000 \AA
\ with a resolution of $\sim$6 \AA. Exposure times ranged from 0.75 --- 1.5 hours.
For galaxies with SDSS r$_{petro} < 20.5$ and r$_{fiber} < 21.5$ , we can observe in gray (and even some bright) time because the galaxies have relatively high surface brightness, the Hectospec fibers are small, and the strip we chose is nearly always far from the moon.

We reduced the data with the standard Hectospec pipeline
(Mink et al. 2007) and derived redshifts with RVSAO (Kurtz \& Mink 1998) with
templates constructed for this purpose (Fabricant et al. 2005). 
Repeat  observations yield robust estimates of the median error  in $cz$ where $z$ is the redshift and $c$ is the speed of light. For emission line objects, the median error (normalized by $(1 + z)$) is 27 km s$^{-1}$; the median for absorption line objects
(again normalized by $(1 + z$)) is 37 km s$^{-1}$
(Geller et al. 2011; Fabricant et al.  2005).

The median redshift of HectoMAP is $ z = 0.34$, $\sim 3.8$ times the depth of the Sloan redshift survey. If the HectoMAP region cuts through great walls at $0.2 < z < 0.6$, they could easily exceed the extent of the Sloan Great wall and still be contained within the survey boundaries. 
Figure \ref{fig:hectomap} is a cone diagram showing the current status of HectoMAP. 

It is always fascinating to see the way each individual Hectospec field contributes to the definition of large-scale structure in the 
survey region. Michael Kurtz and Scott Kenyon made the movie that shows the
data in Figure \ref{fig:hectomap} in the order of acquisition. In the movie, each yellow dot represents a galaxy. The gray arcs are spaced by 0.1 in redshift beginning at $z = 0.1$. Figure \ref{fig:hectomap} shows the final frame of the movie along with an inset of the first slice of the CfA survey from Figure \ref{fig:firstslice} on the same scale.

There are  42,147 redshifts in the current HectoMAP sample (7343 are from the SDSS; the rest are new Hectospec data). The movie gives an idea of the excitement we feel as we watch the survey develop. In contrast with the first slice of the CfA survey where the striking pattern was a surprise, we now expect to see well-delineated, large voids surrounded by the galaxies that define the cosmic web. Like other surveys to its depth, HectoMAP will provide constraints on the characteristics of the structure as a function of cosmological epoch.  

Because HectoMAP is a survey of red galaxies, the large-scale features of the map are more sharply defined than they would be if we included the less strongly clustered blue objects
(Davis \& Geller 1976; Zehavi et al. 2011). The much larger BOSS survey (White et al. 2011) geared toward exploitation of the baryon acoustic peak, select even redder objects essentially eliminating objects with $z \lesssim 0.45$. The BOSS survey is optimized to take advantage of a sparsely sampled large volume at greater redshift than the range we sample with HectoMAP. In contrast,  
the dense sampling of HectoMAP enables a focus on clusters of galaxies and their relationship with large-sale structure.

Based on our earlier SHELS survey (Geller et al. 2010; Kurtz et al. 2011), HectoMAP should contain $\sim 50$ clusters with rest-frame line-of-sight velocity dispersions $\gtrsim 600$ km s$^{-1}$. The survey should yield a robust determination of the cluster mass function in the range $0.2 < z < 0.55$. Systems of galaxies are evident throughout this range in Figure \ref{fig:hectomap}; it is, of course, harder to see them by eye at greater redshift because the line-of-sight velocity dispersion is a small fraction of the mean redshift. The set of systems extracted from the full survey promise, among many other applications,  a strong test of cluster catalogs based on
identification of the red sequence in photometric data. 

As an example, Figure \ref{fig:gmbcg} shows a small test of the GMBCG catalog (Hao et al. 2010). In this catalog, clusters are selected from the SDSS photometric catalog with an algorithm based on identifying  the red sequence and the brightest cluster member.
The figure shows a sample of the highest confidence GMBCG clusters (with at least 15 members) that lie within the most complete HectoMAP region right ascension range $13.33^h < \alpha_{2000} < 15.33^h$.  The yellow bar is centered at the mean photometric redshift for each GMBCG cluster in the range of this portion of HectoMAP and the bar extends for 
$\pm \Delta{z}/(1+z) \sim 0.01$. The black points are galaxies in the HectoMAP survey; these red galaxies are the ones that populate a typical cluster red sequence. 

The correspondence between the redshift survey and this cluster catalog is poor. Many of the GMBCG clusters do not exist at all or they are superpositions of cuts through the cosmic web.
Some of the systems which appear as clean ``fingers'' in HectoMAP  are  missed by the GMBCG scheme. The discrepancies underscore the importance of spectroscopy for understanding structure in the universe and its evolution. As we complete HectoMAP we will extend this test; it is reminiscent of the days of testing the  Abell (Abell, Corwin \& Olowin 1989) catalog 
where many of the least rich systems are also superpositions.

Perhaps the most striking features of the HectoMAP display are the structures that appear to cross the survey roughly perpendicular to the line-of-sight at several redshifts. Because the survey is incomplete, we cannot apply an objective measure of these structures (and features in other orientations) at this stage. 
Characteristics of the HectoMAP walls include the presence of many systems of galaxies embedded in the walls. 

There is an impressive structure crossing the entire right ascension range at $z\sim 0.3$; the 
comoving length of this structure is $\sim 500h^{-1}$ Mpc compared with 315$h^{-1}$ Mpc for the the Sloan Great Wall (Gott et al. 2005).
The fingers corresponding to systems are clear in this HectoMAP wall at z = 0.3. At $z \sim 0.5$ there is another structure which appears likely to cross the whole survey. Gaps in the HectoMAP coverage are more obvious at greater redshift because we have not reached the survey magnitude limit
throughout the right ascension range. 
Of course, HectoMAP is a narrow slice and thus we do not yet have a constraint on the extent of this structure in declination. However, it is unlikely that one-dimensional structures would just happen to lie in this narrow slice. 
In the next year we will see whether these HectoMAP structures, like the Sloan Great Wall, are a possible challenge to current wisdom.

\section {Concluding Remarks}

\begin{figure}[htb]
\centerline{\includegraphics[width=6.5in]{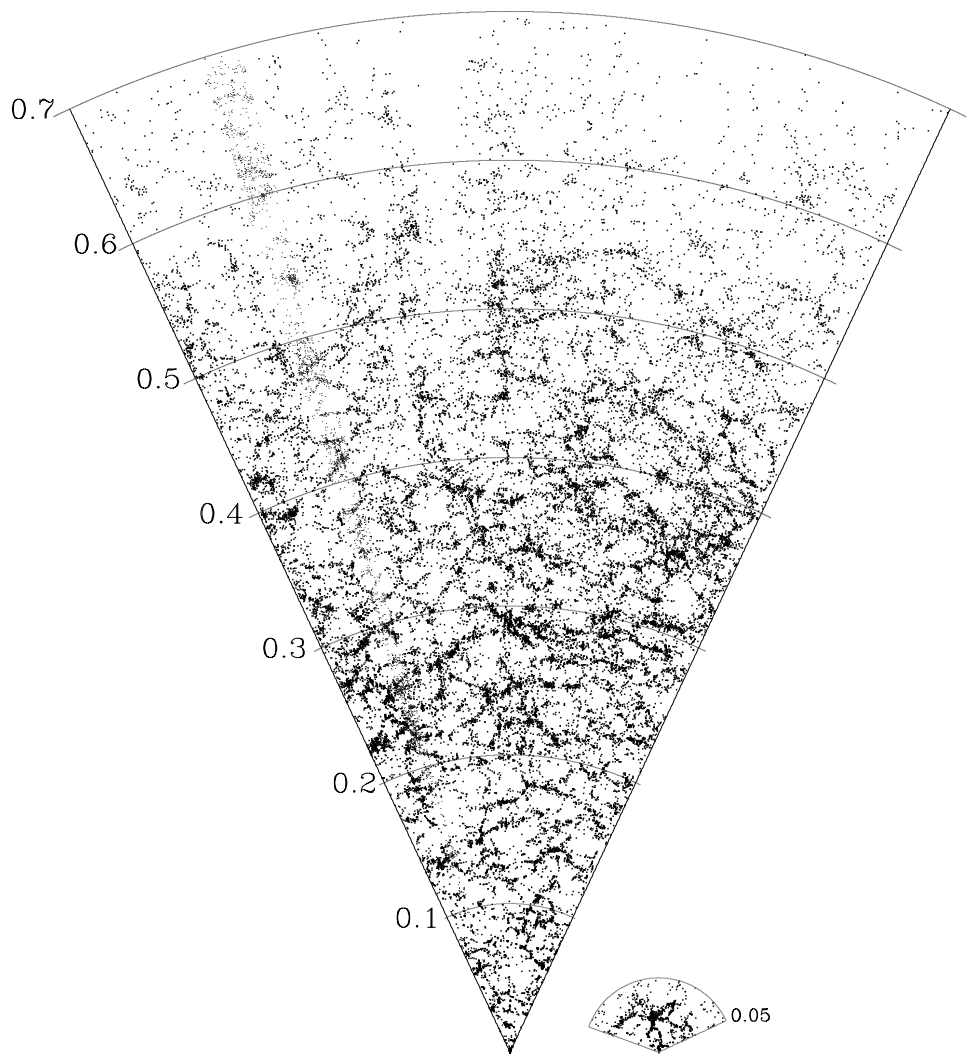}}
\vskip 5ex
\caption{Cone diagram for the current status of the HectoMAP redshift survey. Note the
apparent walls crossing the entire survey and the many well defined large voids. The inset shows 
Figure \ref{fig:firstslice} on the same scale. The radial scale is the redshift. The right ascension range of HectoMAP is $13.33^h < \alpha_{2000} < 16.67^h$. Note the structures crossing the entire survey at $z \sim 0.3$ and $z \sim 0.5$. The on-line journal article includes a 
video showing the development of the map as the data are acquired; the cone diagram shown in this figure is the last frame of the video.
\label{fig:hectomap}}

\end{figure}

\begin{figure}[htb]
\centerline{\includegraphics[width=6.5in]{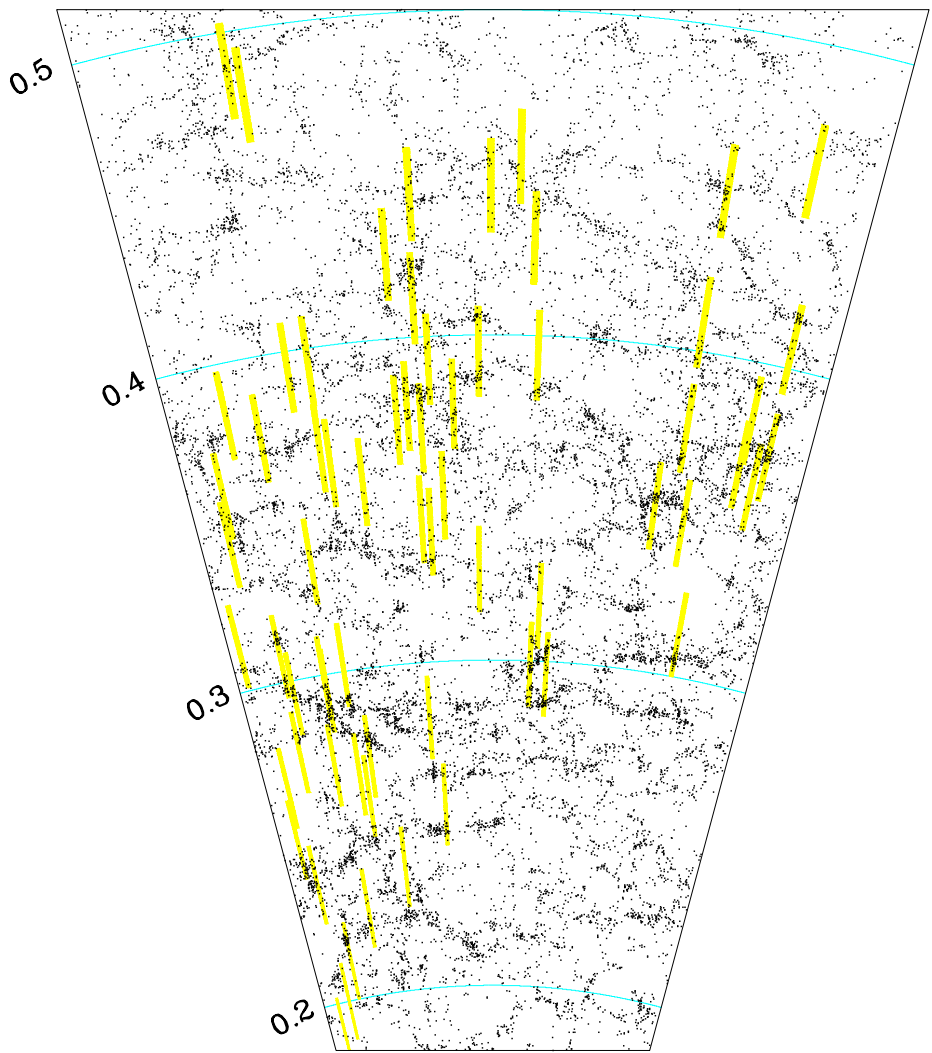}}
\vskip 5ex
\caption{Comparison between a portion of the HectoMAP redshift survey and the GMBCG cluster catalog based on identification of the red sequence. The radial coordinate is the redshift; the right ascension range is $13.33^h < \alpha_{2000} < 15.33^h$. Points represent galaxies in the HectoMAP redshift survey.
Yellow bars are centered at the positions of the GMBCG clusters; the length of bar is the error in the photometric redshift estimate. 
\label{fig:gmbcg}}

\end{figure}

Comparison of the first CfA slice and HectoMAP (Figure \ref{fig:hectomap}) is a visual demonstration that we have come a long way since 1986. From measuring 25 redshifts per night (one at a time) with a 1.5-meter telescope, we have progressed to measuring 2000 per night with a wide-field instrument on a 6.5-meter telescope. Our first steps seem remarkably small compared with  the reach of modern instruments. Nonetheless those first steps were big enough to reveal the basic architecture of the universe.

With the stunning increase in the amount and quality of data over the past decades, maps of the universe cover more of the sky to greater and greater depth. Ambitious proposals for new wide-field instruments with ever greater 
power show that there is no sign of loss of interest in extending these maps.

It has been an extraordinary, heady experience to be a part of the discovery of the largest patterns we know in nature. As part of  that experience I have often tried to put myself in Charles Kittel's shoes when he gave me the advice that so changed the course of my career.

In spite of the advances in our ability to observe the universe, truly fundamental puzzles remain. Dark matter has been with us since its discovery by Zwicky in 1933, and we {\it still} have no idea what it is. Even worse,  the equally mysterious dark energy dominates the energy density of the universe. We blithely say that we have a standard model when we really have no fundamental understanding of its main ingredients. There is certainly still room for 
revolution here.

On the other hand, the structure of our field is becoming so rigidly dominated by large groups that it is hard for someone with truly original, but solid  ideas to get a position with the freedom to pursue a distinctive course. I would encourage senior people to give some thought to the structure they have set up that so discourages risk-taking. We all need to  ask ourselves some hard questions about the kinds of opportunities we provide in science and the way we evaluate the most creative young scientists.

The universe is staggeringly vast. Just as the large-scale patterns in the universe were a surprise, there will be more surprises. Some of these surprises will, I hope, still be in work
done by imaginative individuals looking in neglected corners where important questions can be found and answered.  

\acknowledgments

No one makes a scientific career alone. I have had the privilege of supporting the careers of 
talented men and women who chose to have me supervise their PhD theses and they, in turn, have supported me. I have had extraordinary collaborators throughout my career and  I have had generous support from people I admire.

The new work reported here will be published in full elsewhere. In advance, I thank Dan Fabricant, Scott Kenyon, Ken Rines, and Warren Brown who have all participated in these projects and have provided advice and counsel whenever needed. Perry Berlind and Mike Calkins operated the Hectospec
to acquire the data displayed here. Susan Tokarz reduced the data. We thank Ken Rines for providing HeCS data in advance of formal publication.

The Smithsonian Institution supports the research of Margaret Geller and Michael Kurtz, and 
during 2010-2011, a Smithsonian Fellowship partially supported Antonaldo Diaferio.
INFN grant PD51 and the PRIN-MIUR-2008 grant 2008NR3EBK also support Antonaldo Diaferio.

\clearpage

\end{document}